\documentclass[10 pt, letter]{article}
\usepackage{graphicx}
\usepackage{caption}
\usepackage{amsmath}
\usepackage{amsthm}
\usepackage{enumerate}
\usepackage{subfig}
\usepackage{float}
\usepackage{amsfonts}
\usepackage{amssymb}
\usepackage{anysize}
\usepackage{helvet}
\usepackage{setspace}
\usepackage{multirow}
\usepackage{tabularx}
\usepackage{tabulary}
\usepackage{longtable}
\usepackage{float}
\usepackage{mathtools}
\title{\bf $C^{12}-C^{13}$ Quantum computer with longitudinal rf-magnetic field}
\author{\Large  Gustavo  V. L\'opez\footnote{gulopez@udgserv.cencar.udg.mx}
 and Jorge Lizarraga\footnote{jorge.a.lizarraga.b@gmail.com}~\\ \\
 Departamento de F\'{i}sica, CUCEI, \\ Universidad de Guadalajara,\\
 Blvd. Marcelino Garc\'{i}a Barragan y Calzada Ol\'{i}mpica, \\Ê44200 Guadalajara, Jalisco, Mexico}
\begin{document}
\maketitle
\noindent
{\bf Keywords:} Quantum computer, diamond, universal quantum gates\\Ê
\begin{abstract}
\noindent
It is shown that in the diamond $C^{12}$ structure with a linear chain of $C^{13}$ atoms which is inside a transverse static field,
having a gradient along the linear chain  and an rf-magnetic field in a plane with a component in the direction of the static field,
one qubit rotation, the Controlled-Not (CNOT),  the Controlled-Controlled-NoT (CCNOT) quantum gates, and teleportation algorithm 
can be implemented using integer multiples of $\pi$-pulses.   
\end{abstract}
\newpage
\section{ Introduction}
Despite of the great experimental and theoretical efforts, quantum computer realization for 1000 qubits remains being the goal of computer science and physical science. The two main problem
faced on this goal are technological limitations [(for example number of controlled ions trapped, maximum number of signal that can be register in NMR system), and the decoherence due to interaction 
of any quantum system with the environment \cite{Warren,Vander,Hol,MonKim}. Several quantum systems have been proposed for quantum computation \cite{Walther,Jak,Youn,Chi,Kit,Child,Ber1}, an some of the attractive (due to their high decoherence time \cite{Wata})  are nuclear 
spin systems. One resent proposal of this type \cite{GV1} is the $C^{12}-C^{13}$ diamond structure for a possible quantum computer, where a chain of $C^{13}$ stable isotopes of spin one half are embedded       
periodically along a given line of the diamond lattice. A transverse static field with some gradient along this line, and transversally to it,  and rf-magnetic field  are applied. The universal CNOT and CCNOT 
quantum gates are obtained with a single $\pi$-pulses (evolution time is $\tau=\pi/\Omega$, being $\Omega$ the Rabi's frequency), and Hadamar's gate is obtained with a $\pi/2$-pulse (evolution time of
the quantum system is $\tau=\pi/2\Omega$). In this paper, we show that even with an rf-magnetic field having a component along the direction of the static magnetic field, it is possible to obtain the 
CNOT, CCNOT universal quantum gates with a $2\pi$-pulse, the Hadamar's gate with a $\pi$-pulse. In addition, the implementation of teleportation algorithm using three qubits is presented with its value
of goodness (different of Fidelity \cite{Fuc}) of the reproduction of the algorithm and its associated Boltzmann-Shanon's entropy \cite{Bolz,Sha}.  
\begin{figure}[h]
\includegraphics[scale=0.6,angle=270]{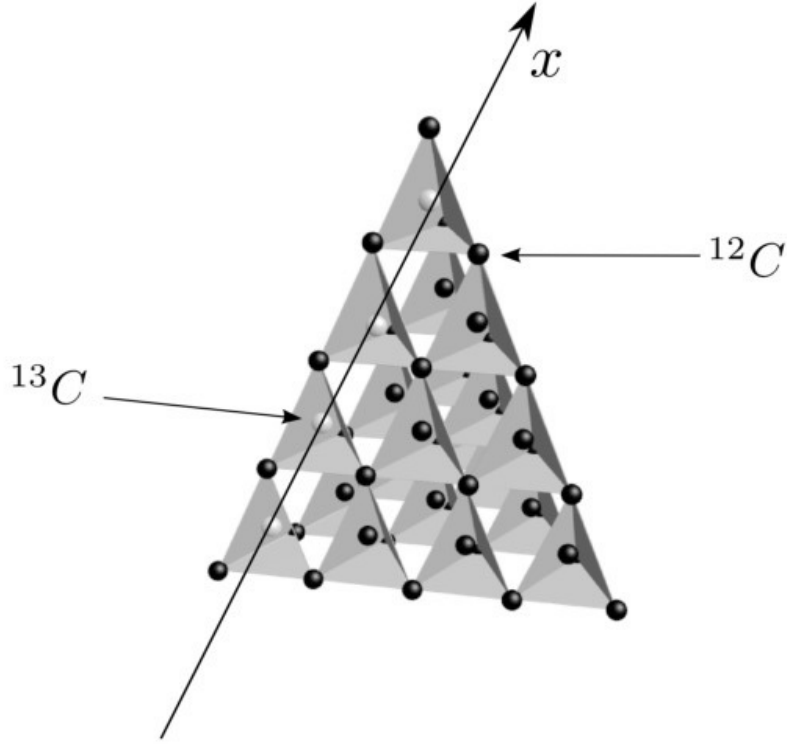}
    \caption{Diamond $C^{12}$-$C^{13}$. }
    \label{fig:A}
\end{figure}
\section{Analytical Approach}
The diamond structure  $C^{12}$ (spin zero) with the line of isotopes $C^{13}$ (spin one half) along the x-axis is shown in Figure 1. The spin-spin interaction is due to their dipole magnetic moments 
interaction \cite{Jack}
\begin{equation}
U=\frac{\mu_0}{4\pi}\frac{3({\bf m}_1\cdot {\bf x})({\bf m}_2\cdot{\bf x})-{\bf m}_1\cdot{\bf m}_2}{|{\bf x}|^3},
\end{equation}  
where $|{\bf x}|$ is the distance separation between the magnetic moments ${\bf m}_1$ and ${\bf m}_2$ which are related with the nuclear spin as
\begin{equation}
{\bf m}_i=\gamma {\bf S}_i, \quad i=1,2.
\end{equation}
being $\gamma$ the gyromagnetic ratio ($\gamma\approx 2.675\times 10^8rad/T\cdot s$). Assuming Ising interaction, this energy can be written as
\begin{equation}
U=\frac{2J}{\hbar}S_1^zS_2^z, \quad\quad J=\frac{\mu_0\gamma^2\hbar}{4\pi a^3},
\end{equation}
where $\hbar$ is the Planck's constant divided by $2\pi$ ($\hbar\approx 1,054571818 \times 10^{ -34}J\cdot s$) and $a=|{\bf x}|\sim 10^{-10}m$ is the $C^{13}-C^{13}$ nuclear separation in the diamond crystal. 
The propose magnetic field is
\begin{equation}
{\bf B}(x,t)=\bigl(0, b\cos\theta,-b\cos\theta+B_0(x)\bigr), \quad\quad \theta=\omega t+\varphi,
\end{equation}
where $\omega$, $\varphi$ and $b$ are the frequency, the phase and the amplitude of the rf-magnetic field, $B_0(x)$ is the static magnetic field transverse to the line of $C^{13}$'s. The interaction of the $C^{13}$'s with the magnetic field is given by\cite{Jack} 
\begin{equation}
U=-\sum_{i=1}^n{\bf m}_i\cdot{\bf B}_i(t), \quad\quad {\bf B}_i(t)={\bf B}(x_i,t),
\end{equation}
where $x_i$ is the position of the ith-atoms $C^{13}$. The Hamiltonian of the system  considering Ising interaction at first and second neighbor is
\begin{equation}
\widehat{H}=U+\frac{2J}{\hbar}\sum_{i=1}^{n-1}S_i^zS_{i+1}^z+\frac{2J'}{\hbar}\sum_{i=1}^{n-2}S_i^zS_{i+2}^z. 
\end{equation}
After rearranging terms, this Hamiltonian can be written as
\begin{equation}
\widehat H=\widehat {H}_0+\widehat W,
\end{equation}
where the operators $\widehat{H}_0$ and $\widehat W$ are defined as
\begin{equation}
\widehat{H}_0=-\sum_{i=1}^n\omega_iS_i^z+\frac{2J}{\hbar}\sum_{i=1}^{n-1}S_i^zS_{i+1}^z+\frac{2j'}{\hbar}\sum_{i=1}^{n-2}S_i^zS_{i+2}^z
\end{equation}
and
\begin{equation}
\widehat{W}=\Omega\sum_k^n\biggl\{\frac{1}{2i}(S_k^+-S_k^{-})\cos\theta+S_k^z\sin\theta\biggr\},
\end{equation}
being $\omega_i=\gamma B_0(x_i)$ the Larmore's frequency associated to the ith-atoms $C^{13}$,  $\Omega=\gamma b$ is the so called Rabi's frequency, and $S_k^+$ and $S_k^{-}$ are the ascend and descend 
operators, $S_k^{\pm}=S_k^x\pm S_k^y$. The n-qubits registers $\{|\xi\rangle=|\xi_n,\dots,\xi_1\rangle\}$ ($\xi_i=0,1$ for digital notation, and $\xi=1,\dots,2^n$ for decimal notation) form the basis for the Hilbert space
of dimensionality $2^n$, our qubit is just the spin one half of the nucleus of the atom $C^{13}$. Now, having the known operations
\begin{equation}
S_i^{z}|\xi_n,\dots,\xi_i,\dots,\xi_1\rangle=\frac{(-1)^{\xi_i}\hbar}{2}|\xi_n,\dots,\xi_i,\dots,\xi_1\rangle,
\end{equation}
\begin{equation}
S_i^{+}|\xi_n,\dots,\xi_i,\dots,\xi_1\rangle=\hbar\delta_{\xi_i,0}|\xi_n,\dots,\xi_i+1,\dots,\xi_1\rangle,
\end{equation}
and 
\begin{equation}
S_i^{-}|\xi_n,\dots,\xi_i,\dots,\xi_1\rangle=\hbar\delta_{\xi_i,1}|\xi_n,\dots,\xi_i-1,\dots,\xi_1\rangle,
\end{equation}
being $\delta_{ij}$ the Kronecker's delta. The solution of the eigenvalue problems
\begin{equation}
\widehat{H}_0|\xi\rangle=E_{\xi}|\xi\rangle 
\end{equation} 
is given by the eigenvalues
\begin{equation}
E_{\xi}=-\frac{\hbar}{2}\sum_{i=1}^n\omega_i+\frac{\hbar}{2} J\sum_{i=1}^{n-1}(-1)^{\xi_i+\xi_{i+1}}+\frac{\hbar}{2} J'\sum_{i=1}^{n+2}(-1)^{\xi_i+\xi_{i+2}},
\end{equation}
and the eigenfunctions are just $\{|\xi\rangle\}$, for $\xi=1,\dots,2^{n}$. Therefore, proposing a solution of the Schr\"odinger's equation
\begin{equation}
i\hbar\frac{\partial |\Psi\rangle}{\partial t}=\widehat{H}|\Psi\rangle,
\end{equation}
of the form
\begin{equation}
|\Psi(t)\rangle=\sum_{\eta=1}^{2^n}e^{-iE_{\eta} t/\hbar} D_{\eta}(t)|\eta\rangle, 
\end{equation}
we get the equation for the coefficients $D_{\xi}(t)$ as
\begin{equation}\label{eqD}
i\dot{D}_{\xi}(t)=-i\frac{\Omega\cos\theta}{2}\sum_{\eta=1}^{2^n}\sum_{k=1}^ne^{i\omega_{\xi\eta} t}D_{\eta}(t)\langle \xi|S_k^{+}-S_{k}^{-}|\eta\rangle+\frac{\Omega (-1)^{\xi_k}}{2}D_{\xi}(t)\sin\theta,
\end{equation}
where the frequency $\omega_{\xi\eta}$ has been defined as
\begin{equation}
\omega_{\xi\eta}=\frac{E_{\xi}-E_{\eta}}{\hbar}.
\end{equation}
We can make all system without units by defining the new time evolution as
\begin{equation}\label{tau}
\tau=\omega_0 t, \quad\quad \omega_0=2\pi kHz.
\end{equation}
Then the constant $\Omega/\omega_0$, $J/\omega_0$ and $\omega_i/\omega_0$ become without units, but the equation (\ref{eqD}) would be exactly of the same form. So, from now on,
we will talk about the time evolution in terms of the variable (\ref{tau}), and all the constant must be though as its value times $2\pi kHz$. 
\section{Quantum Universal Gates}
Equations (\ref{eqD}) are solved numerically using Runge-Kutta method at fourth order. We made the simulation of the NOT quantum gate (1-qubit), Controlled-Not gate (2-qubits) and 
Controlled-Controlled-Not gate (3-qubits)  to show the realization of a quantum computer with this magnetic field configuration in the $C^{12}-C^{13}$ diamond structure. The parameters
used in our simulations are given by
\begin{subequations}
\begin{eqnarray}
& & 1-qubit  : \omega_1=100, \quad\Omega=0.1\\
& & 2-qubits: \omega_1=100, \quad \omega_2=200,\quad \Omega=0.1,\quad J=2\\
& & 3-qubits: \omega_1=100, \quad\omega_2=200,\quad \Omega_3=300, \quad\Omega=0.1,\quad J=2,\quad J'=1/4,
\end{eqnarray}    
\end{subequations}
and the initial conditions are
\begin{subequations}
\begin{eqnarray}
& & 1-qubit  :  D_1(0)=1,\quad D_2(0)=0\\
& & 2-qubits:  D_1(0)=\frac{1}{\sqrt{16}},\quad D_2(0)=\sqrt{\frac{2}{16}},\quad D_3(0)=\sqrt{\frac{5}{16}}, \quad D_4(0)=\sqrt{\frac{8}{16}}\\
& & 3-qubits:  D_1(0)=D_2(0)=\sqrt{\frac{1}{8}},\quad D_3(0)=D_4(0)=0,\quad D_5(0)=\sqrt{\frac{3}{8}},\quad D_7(0)=\sqrt{\frac{2}{8}},\quad D_8(0)=\frac{1}{\sqrt{8}}\nonumber\\
\end{eqnarray}    
\end{subequations}
Figure 2 shows the probabilities $|D_i(\tau)|^2$ for the system to be in the state $|i\rangle$ as a function of $\tau$. This curves show the 
realization of the quantum gates NOT (transition: $|0\rangle\longleftrightarrow |1\rangle$), CNOT (transition: $|10\rangle\longleftrightarrow |11\rangle$), and CCNOT (transition: 
$|110\rangle\longleftrightarrow |111\rangle$) using a $2\pi$-pulse ($\tau=2\pi/\Omega$). The probabilities of the other states (CNOT and CCNOT) remain constant within and 
error of the order $10^{-5}$. In addition, 
We must observe that a superposition of states (Hadamar quantum gate) is gotten for a time evolution of
$\tau=\pi/\Omega$  ($\pi$-pulse), that is the states of the form
\begin{eqnarray}
& & \frac{1}{\sqrt{2}}\bigl(|0\rangle+|1\rangle\bigr)\quad\quad\hbox{(1-qubit)}
\end{eqnarray} 
where some phases could be as factors on each state. We want to mention that if the initial conditions for CNOT and CCNOT are $D_3(0)=1$ and $D_7(0)=1$ and all other coefficients
are zero, the superposition of states $|10\rangle+|11\rangle$ and $|110\rangle +|111\rangle$ are obtained too.
\begin{figure}[h]
\includegraphics[scale=0.6,angle=0]{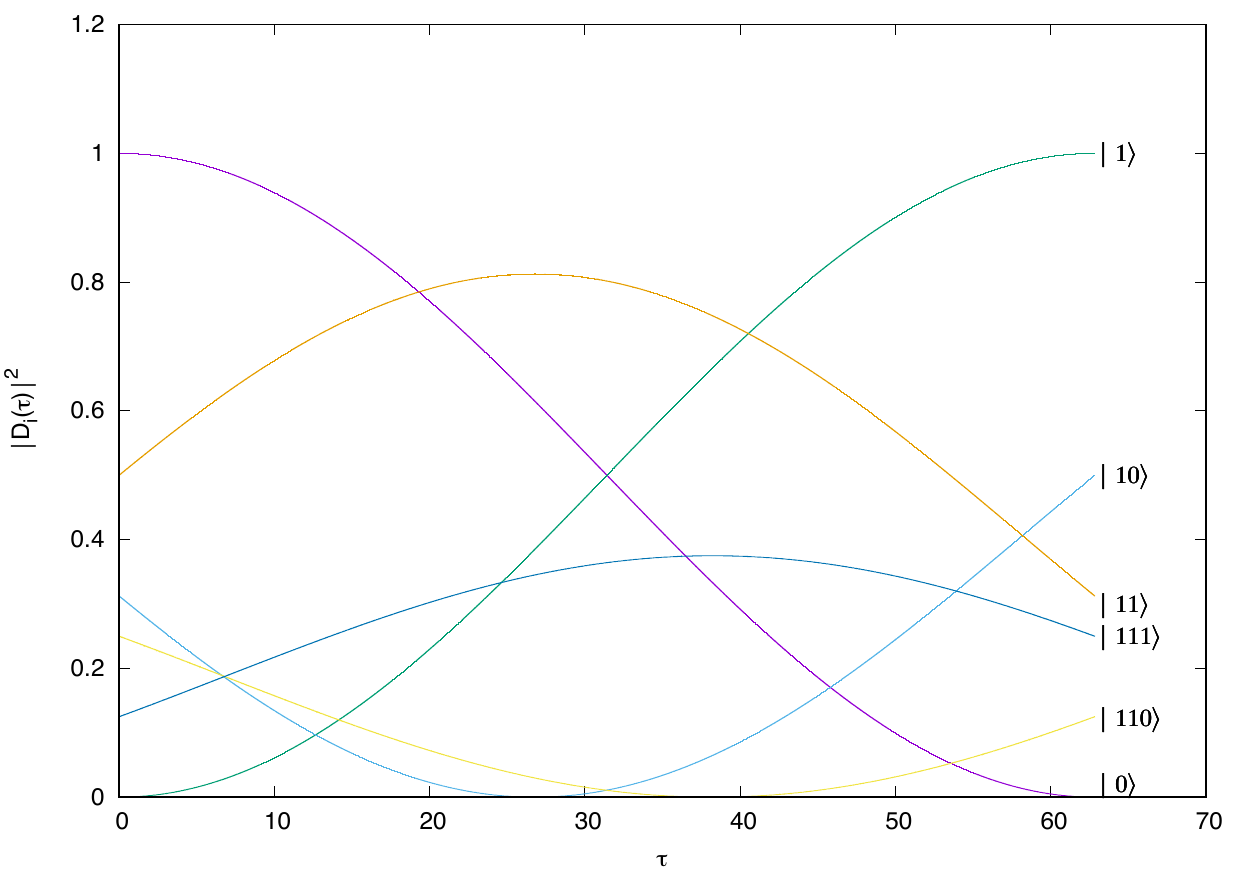}
\centering
    \caption{Transitions: $|0\rangle\longleftrightarrow |1\rangle,\quad |01\rangle\longleftrightarrow |11\rangle,\quad |110\rangle\longleftrightarrow |111\rangle$ }
    \label{fig:A}
\end{figure}
\section{Realization of teleportation algorithm}
Quantum teleportation \cite{Nie}  is one of the most fundamental element in quantum computation and quantum information which deserves its realization for any
quantum computer proposal. Therefore, this algorithm will be studied in this section with three qubits configuration, where the first state will represent Bob, the 
second state will be Alice and the third state will be the teleported state. From the Alice location, we want to transport this state (in general, an unknown state even by Alice) to Bob place.  
If the state we want to transport is
\begin{equation}
|\psi\rangle=c_1|0\rangle+c_2|1\rangle, \quad\quad c_1=\sqrt{\frac{3}{8}},\quad c_2=\sqrt{\frac{5}{8}},
\end{equation}  
the initial state of our 3-qubits system is then
\begin{equation}\label{st0}
|\Psi_0\rangle=|\psi\rangle\otimes|0\rangle\otimes|0\rangle=c_1|000\rangle+c_2|100\rangle.
\end{equation}
The "ideal" (mathematical) teleportation algorithm is represented by quantum circuit  shown on Figure 3 (going from left to right), 
\begin{figure}[h]
\includegraphics[scale=2.0,angle=0]{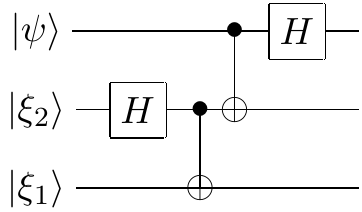}
\centering
    \caption{Teleportation quantum circuit}
    \label{fig:A}
\end{figure}
or written in term of operators  we have
\begin{equation}\label{opT}
T=\bigl(H_3\cdot CN_{32}\bigr)\cdot\bigl(CN_{21}\cdot H_2\bigr),
\end{equation}
where $CN_{ij}$ is the CNOT operation between the qubit "i" and the cubit "j" ($CN_{ij}|\xi_n,\dots,\xi_i,\dots,\xi_j,\dots,\xi_1\rangle=|\xi_n,\dots,\xi_i,\dots,\xi_j\oplus\xi_i,\dots,\xi_1\rangle$), and $H$ is the
 Hadamar operator (1-qubit). The first term,
$\bigl(CN_{21}\cdot H_2\bigr)$, makes the entanglement between Alice(second qubit) and Bob (first qubit), and the term $\bigl(H_3\cdot CN_{32}\bigr)$ represents the Alice operations to produce the teleportation.  The result of
the application of the operator (\ref{opT}) to the initial state (\ref{st0}) can be written as
\begin{equation}
T|\Psi_0\rangle=\frac{1}{2}
\biggl\{|00\rangle\otimes|\psi\rangle+|01\rangle\otimes N_3|\psi\rangle+|10\rangle\otimes Z_3|\psi\rangle+|11\rangle\otimes N_3Z_3|\psi\rangle\biggr\},
\end{equation}
where $N_3$ (NOT gate),  $Z_3$ and $N_3Z_3$ act in the following way
\begin{equation}
N_3|\psi\rangle=c_1|\rangle+c_2|0\rangle,\quad\quad Z_3|\psi\rangle=c_1|0\rangle-c_1|1\rangle, \quad\quad N_3Z_3|\psi\rangle=c_2|0\rangle-c_1|1\rangle. 
\end{equation}
Thus, the state $|\psi\rangle$ has been teleported to Bob's place, and this has occurred instantly (independently how separated are Alice and Bob). Of course, Alice needs to notify (at the speed of light or less) to Bob
about the operation he needs to do in order to get the state $|\psi\rangle$ ($N_3^2=Z_3^2=I$, the identity operator). Notice then that at the end of the algorithm, we have four states with a probability $|c_1|^2/4$ and 
four states with probability $|c_2|^2/4$.  However, in our non-ideal system, we need to apply the correct type of pulse with the desired frequency transition in order to get the state what we want. Let us denote by
\begin{equation}
R_{ij}(\tau),
\end{equation}
the pulse of length $\tau$ done in the system with a resonant frequency $\omega=(E_i-E_j)/\hbar$. Our non-ideal teleportation operator is given by
\begin{equation}
\widetilde{T}=\biggl[R_{26}(\pi)R_{37}(\pi)R_{78}(\pi)R_{15}(\pi)R_{78}(2\pi)R_{57}(2\pi)\biggr]
\biggl[R_{78}(2\pi)R_{34}(2\pi)R_{57}(\pi)R_{13}(\pi)\biggr],
\end{equation}
which represents $\pi$ and $2\pi$ pulses where the Rabi's frequency has been omitted. The last square parenthesis represents the entanglement, and the other one represents the teleportation. The result of this application 
to the initial state (\ref{st0}) can be seen Figure 4, where we have gotten four states with the probability $3/32$ and four states with  probability $5/32$.    
\begin{figure}[h]
\includegraphics[scale=0.8,angle=0]{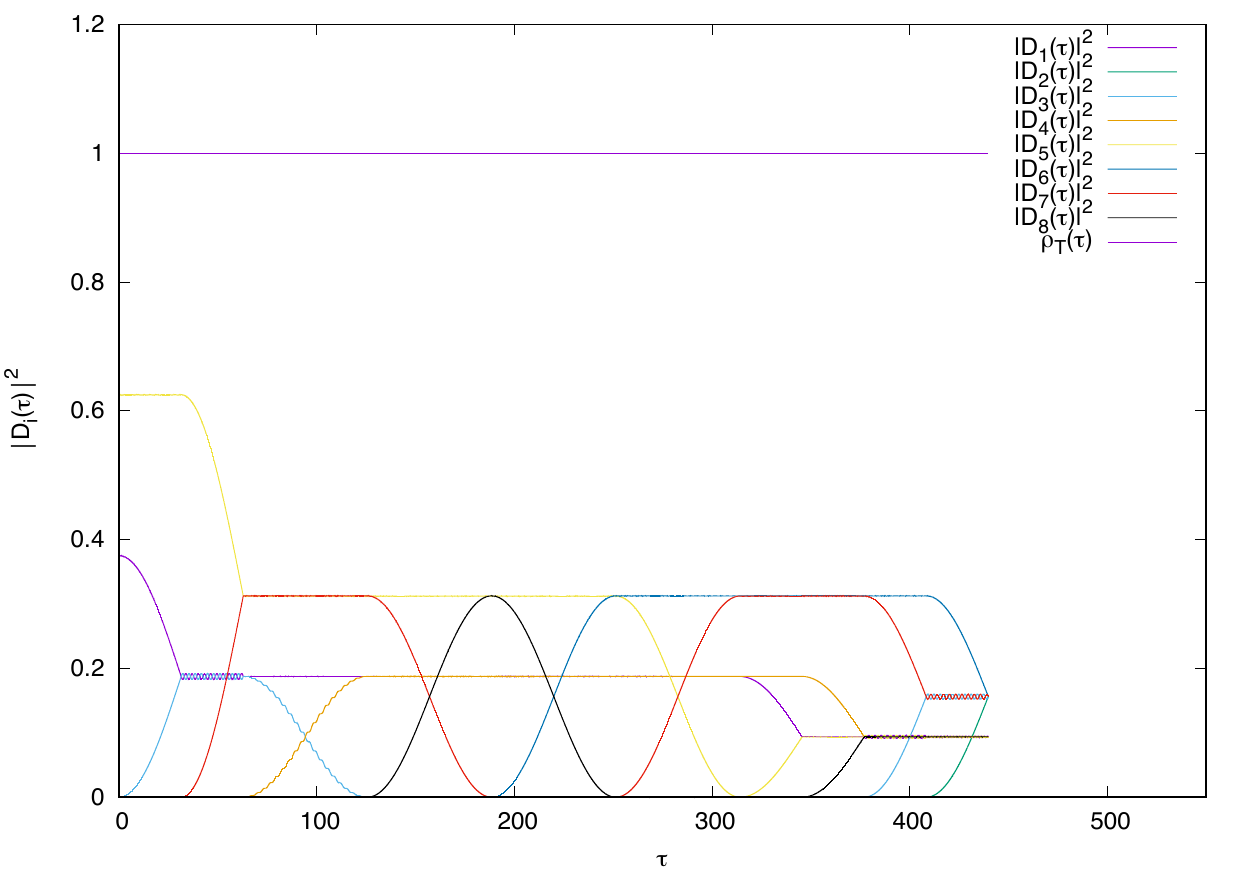}
\centering
    \caption{Teleportation algorithm and total probability $P_t(\tau)$}
    \label{fig:A}
\end{figure}
\newpage\noindent
The goodness of the reproducibility of the teleportation algorithm can be measured by comparing the probabilities of the states between the real case with the ideal case 
at the end of the algorithm, and this can be done through the quantity
\begin{equation}
G=\sum_{i=1}^8\bigl| |D_i^{(ideal)}|^2-|D_i^{(real)}|^2\bigr|,
\end{equation}
which goes from zero (excellent reproducibility) to one (non reproducibility at all). 
In our case, the value obtained is $G=7.4\times 10^{-4}$, meaning that we have a very good reproduction of the teleportation algorithm. \\Ê\\
On the other hand, it is of interest to see how the information of the system is lost during the teleportation algorithm since starting with 2 states we finish 
with 8 states. Therefore, there must be an increasing on Boltzmann-Shanon entropy and its average value,
\begin{equation}
S(\tau)=-\sum_{i=1}^8|D_i(\tau)|^2\ln|D_i(\tau)|^2, \quad \quad \langle S(T)\rangle=\frac{1}{T}\int_0^TS(\tau) d\tau.
\end{equation}
Of course, the maximum value this entropy can have is went all the probabilities are the same (1/8), $S_{max}=\ln 8\approx 2.08$.  Figure 5 shows the evolution of the entropy and its average value, 
and as we expected,  there is an increasing on the entropy and its average value due to the increasing of the number of states involved in the quantum dynamics of teleportation algorithm. The oscillations
shown on this figure are due to CNOT and CCNOT operations which change the states but does increases  the number of states. The rapidly increasing on the entropy is due to Hadamar's operators which increases 
the number of states involved in the quantum dynamics. 
\begin{figure}[H]
\includegraphics[scale=0.8,angle=0]{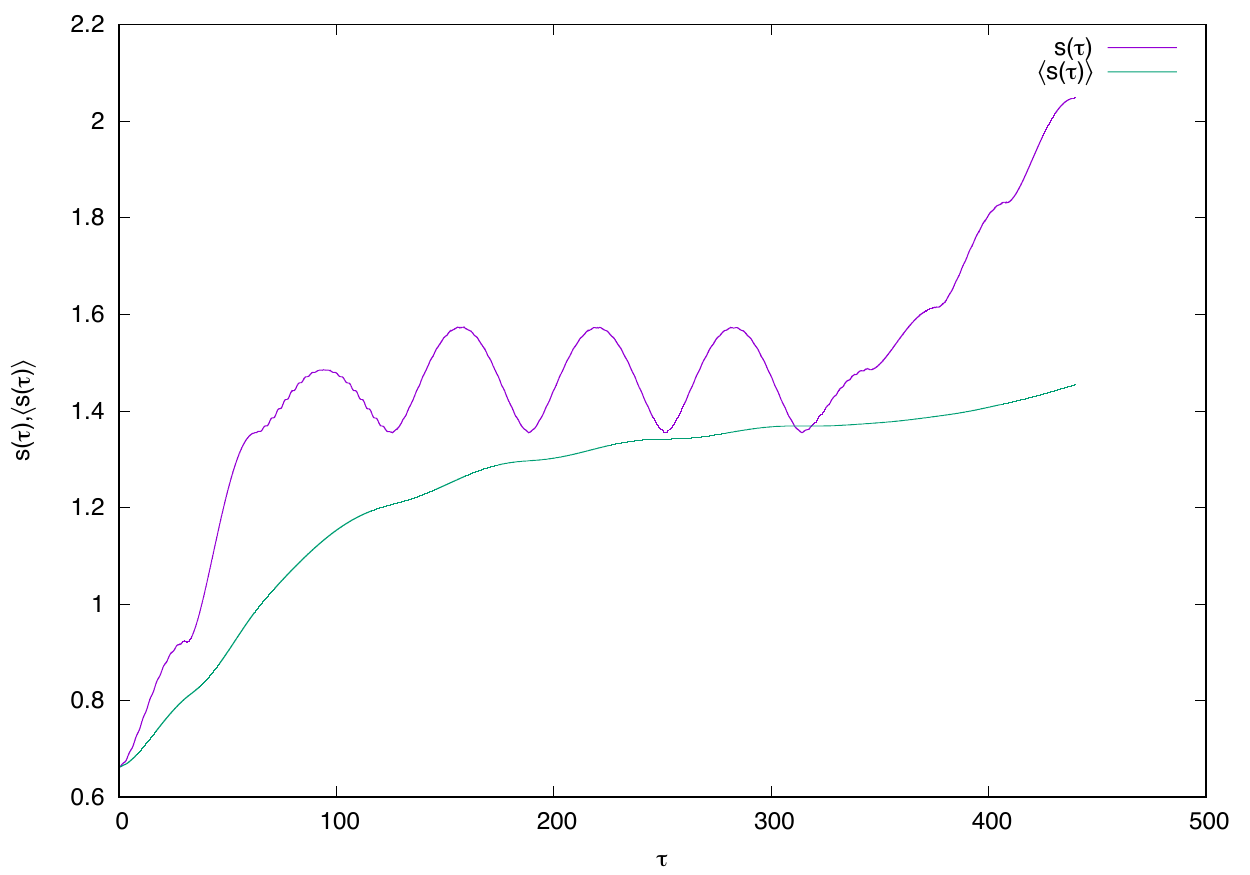}
\centering
    \caption{Boltzman-Shanon entropy for the teleportation algorithm}
    \label{fig:A}
\end{figure}
 \section{Conclusions and Comments}
 We have shown that a solid state quantum computer can be realized using $C^{12}-C^{13}$ structure in a diamond with even a longitudinal rf-magnetic field (one of its component is along the static field).
  The universal CNOT and CCNOT quantum gates and the one qubit rotation were well stablished on the simulations with multiples of a $\pi$-pulse. The teleportation algorithm was also implemented 
  using these quantum gates in a 3-qubits configuration and using 10 pulses, and its associated Boltzmann-Shanon's entropy was determined. We do not think decoherence \cite{ Breuer,CaLeg,UnZu,Paz1,Ven1,Zeh,PazZu,Lind}
   could be of much consideration on this system. However, to make the escalation to many qubits system and the read out \cite{Ru} the final computation result are things to need addressed in the near future.    

\end{document}